\documentclass[pss]{wiley2sp} 
\usepackage{w-greek}         
\usepackage{amsmath}
\usepackage{txfonts}
\usepackage{amsfonts,graphicx}

\usepackage{bm}              
\usepackage{w-greek}         
\usepackage{nicefrac}

\definecolor{Tan}{rgb}{0.737255,0.56078431,0.56078431}
\renewcommand{\copyrightmark}{}
\journalname{Phys. Status Solidi B / \textbf{DOI} 10.1012/pssb.201451174}

\begin{document}

\title{Theoretical electron energy loss spectroscopy of isolated graphene}

\titlerunning{Theoretical electron energy loss spectroscopy of isolated graphene}

\author{Duncan J. Mowbray}

\authorrunning{Duncan J. Mowbray}

\mail{e-mail \textsf{duncan.mowbray@gmail.com}, Phone: +34 943 01 8392, Fax: +34 943 01 8302}
\institute{Nano-Bio Spectroscopy Group and ETSF Scientific Development Center, Departamento de F{\'{\i}}sica de Materiales, Universidad del Pa{\'{\i}}s Vasco UPV/EHU and DIPC, Avenida de Tolosa 72, ES-20018 San Sebasti{\'{a}}n, Spain}


\abstract{A thorough understanding of the electronic structure is a necessary first step for the design of nanoelectronics, chemical/bio-sensors, electrocatalysts, and nanoplasmonics using graphene. As such, theoretical spectroscopic techniques to describe collective excitations of graphene are of fundamental importance. Starting from density functional theory (DFT), linear response time dependent DFT in frequency-reciprocal space within the random phase approximation (TDDFT-RPA) is used to describe the loss function $-\Im\{\varepsilon^{-1}(\textbf{q},\omega)\}$ for isolated graphene.  To ensure any spurious interactions between layers are removed, both a radial cutoff of the Coulomb kernel, and extra vacuum directly at the TDDFT-RPA level are employed.  A combination of both methods is found to provide a correct description of the electron energy loss spectra of isolated graphene, at a significant reduction in computational cost compared to standard methods. 
}

\maketitle  

\section{Introduction}

In the last ten years, graphene \cite{novoselov,Novoselov05N,Geim07NM,CastroNeto09RMP,Novodsel1,Novodsel2,Novodsel3,Novodsel4,Novodsel6,Pletikos,RevModPhys.83.1193,RevModPhys.83.407,RevModPhys.81.109,0034-4885-74-8-082501} has become the favourite playground of researchers for testing methods for modelling the electronic structure of low-dimensional systems \cite{TDDFTGraphene,GrapheneExciton}.  This is because the simplicity of its atomic structure means there is less ambiguity with regards to its electronic structure.  

For this reason, theoretical spectroscopy techniques \cite{Pichler,EELS,splitting,Marinopoulos02PRL,Kramberger08PRL,MiskoSpe,DespojaGraphene,response1,response2} and experimental electron energy loss spectroscopy (EELS) \cite{exp1,2D-exp,exp2,KrambergerGrapheneEELS} have been intensively applied to graphene.  Thus, graphene offers an ideal benchmark for comparing theoretical spectroscopy methods, and there reliability for low-dimensional systems.  Perhaps more importantly, graphene allows one to directly probe a surface without having to separate out the influence of an embedded bulk material.

In this paper, we apply linear response time dependent density functional theory in frequency-reciprocal space within the random phase approximation (TDDFT-RPA) to describe the loss function $-\Im\{\varepsilon^{-1}(\textbf{q},\omega)\}$, absorbance $\Im\{\varepsilon(\textbf{q},\omega)\}$, and dielectric function $\Re\{\varepsilon(\textbf{q},\omega)\}$ of graphene.  Although TDDFT-RPA is a method specifically designed for describing bulk systems, it has recently been increasingly applied to low-dimensional materials (molecules, nanotubes, layers, surfaces, etc.).  This has required a reformulation of TDDFT-RPA to remove spurious interactions between periodic images due to the long-ranged Coulomb interaction.  

One often used technique is the ``radial cutoff'' method \cite{RadialCutoff,KristenDichalcogenides2013}.  Here, the Coulomb interaction between images is explicitly removed by employing a truncated translationally invariant form of the Coulomb interaction.  However, a radial cutoff method becomes cumbersome when describing a bulk surface, where the amount of vacuum required at the DFT level must then be larger than the slab's thickness.  

As a means of overcoming this limitation, one may use the ``zero padding'' technique introduced herein.  With this technique, the unit cell is augmented by additional padding at the TDDFT-RPA level.  By combining these two techniques, one may ensure that all interactions within the supercell are included, and all spurious image--image interactions are removed.

This paper is organized as follows.  In Section~\ref{Methodology} the computational details of the DFT and TDDFT-RPA calculations performed herein are described, followed by a brief review of the TDDFT-RPA formalism, and the radial cutoff and zero padding methodologies.  In Section~\ref{Results} the various TDDFT-RPA techniques employed (standard DFT, zero padding, and radial cutoff) are directly compared with the experimental loss function for graphene; the convergence of graphene's plasmon energies and spectra with the vacuum layer is shown; and the dispersion of the converged loss function, absorbance, and dielectric function of graphene when employing a combination of radial cutoff and zero padding techniques is provided.  This is followed by concluding remarks in Section~\ref{Conclusions}.

\begin{figure*}
\sidecaption
\includegraphics[width=1.41\columnwidth]{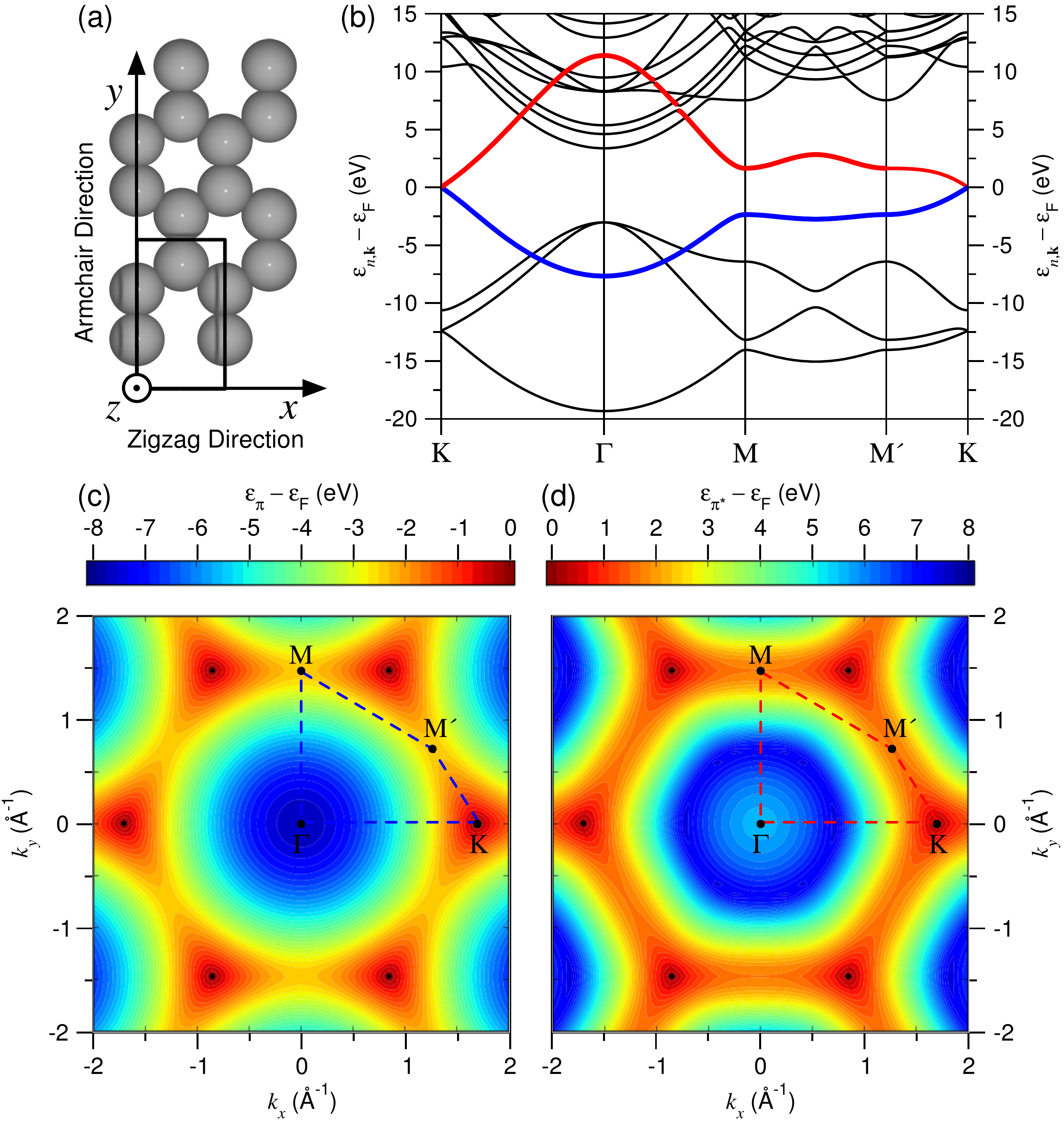}
\caption{(a) Schematic of the orthorhombic graphene unit cell repeated twice in the surface plane.  The $x$-direction corresponds to the zigzag direction or circumference of a zigzag SWNT, while the $y$-direction corresponds to the armchair direction or circumference of an armchair SWNT.  The $z$-direction is normal to the graphene surface. (b) Graphene band structure $\varepsilon_{n,\mathbf{k}}$ in eV relative to the Fermi energy \(\varepsilon_F\) along the high symmetry $\mathrm{K}\rightarrow \Gamma \rightarrow \mathrm{M} \rightarrow \mathrm{M}'\rightarrow\mathrm{K}$ directions. Thick lines are the occupied $\pi$ (blue) and unoccupied $\pi^*$ (red) bands. Fermi surfaces $\varepsilon$ in eV relative to the Fermi level $\varepsilon_F$ for the (c) valence band $\varepsilon_\pi$ and (d) conduction band $\varepsilon_\pi^*$ of graphene, calculated over the Brillouin zone, with reciprocal lattice vectors $k_x$ and $k_y$ in~\AA$^{-1}$ to the zigzag and armchair directions in graphene, respectively.
}\label{FermiSurface}
\end{figure*}

\section{Methodology}\label{Methodology}

All DFT calculations were performed using the real-space
projector augmented wavefunction (PAW) method code \textsc{gpaw} \cite{GPAW,GPAWRev}, with
a grid spacing of 0.2~\AA, and the local density approximation (LDA) \cite{LDA} for the exchange and correlation functional.  An electronic temperature of $k_B T \approx$ 0.05~eV was used to obtain the occupation of the Kohn-Sham (KS) orbitals, with all energies extrapolated to $T = 0$ K, and one unoccupied band per C atom included to improve convergence.

Structural minimization was performed until a maximum force below 0.05~eV/\AA\ was obtained.  An orthorhombic $2.46\times4.26\times L_z$~\AA$^3$\ supercell was employed, where $L_z =$~8, 10, 12, 16, 20, 24, 40, 80, or 160~\AA\ is the length of the unit cell in the $z$-direction.  The supercell consists of four C atoms, as shown schematically in Fig.~\ref{FermiSurface} (a).  Non-periodic boundary conditions were enforced in the $z$-direction normal to the graphene surface, so that both the electron density and KS wavefunctions $\rightarrow $ 0 as $z \rightarrow$ 0 or $z\rightarrow L_z$.   A Monkhorst-Pack $k$-point sampling of 25 $k$-points along the zigzag direction, and 15 $k$-points along the armchair direction of the graphene surface was employed to converge the electron density, yielding a longitudinal momentum transfer resolution $\Delta q$ of 0.102~\AA$^{-1}$ and 0.098~\AA$^{-1}$ respectively.  A finer $75\times45$ $k$-point mesh was employed to obtain a finer  $\Delta q$ of  0.034~\AA$^{-1}$ and 0.033~\AA$^{-1}$ for the calculation of the loss function and dielectric function's dispersion.  At the TDDFT-RPA level, eight unoccupied bands per C atom and 105 \textbf{G}-vectors ($\varepsilon_{\textit{cut}}\approx 36$~eV) were included, which was found to be more than sufficient to converge the loss function for energies up to 50~eV.

Calculations of the dielectric response function have been performed using TDDFT-RPA, as implemented in \textsc{gpaw} \cite{response1,response2}.  Within this framework the Fourier transform of the non-interacting density-density response function $\chi_{\textbf{GG}'}^0(\textbf{q},\omega)$ for momentum transfer $\textbf{q}$ at energy $\hbar\omega$ is given by
\begin{eqnarray}
\chi_{\textbf{G}\textbf{G}'}^0(\textbf{q},\omega) &=& \frac{1}{\Omega}\sum_{\textbf{k}}\sum_{n,n'}\frac{f_{n\textbf{k}} - f_{n'\textbf{k}+\textbf{q}}}{\omega + \varepsilon_{n\textbf{k}} - \varepsilon_{n'\textbf{k}+\textbf{q}} + i\gamma}\nonumber\\
&&\times\int_{\Omega}d\textbf{r}\psi^*_{n\textbf{k}}(\textbf{r})e^{-i(\textbf{q}+\textbf{G})\cdot\textbf{r}}\psi_{n'\textbf{k}+\textbf{q}}(\textbf{r})\nonumber\\
&&\times\int_{\Omega}d\textbf{r}'\psi_{n\textbf{k}}(\textbf{r}')e^{i(\textbf{q}+\textbf{G}')\cdot\textbf{r}'}\psi_{n'\textbf{k}+\textbf{q}}^*(\textbf{r}').\label{chi0}
\end{eqnarray}
Here the sum is over reciprocal lattice vectors $\textbf{k}$ and band numbers $n$ and $n'$, with $\varepsilon_{n\textbf{k}}$ the eigenenergy of the $n^{\textrm{th}}$ band at $\textbf{k}$, $f_{n\textbf{k}}$ the Fermi-Dirac occupation of the $n^{\textrm{th}}$ band at $\textbf{k}$, $\gamma$ the peak broadening, $\Omega$ the volume of the supercell, $\textbf{G}$ and $\textbf{G}'$ the reciprocal unit cell vectors, and $\psi_{n\textbf{k}}(\textbf{r})$ the real-space KS wavefunctions for the $n^{\textrm{th}}$ band with reciprocal lattice-vector $\textbf{k}$.  The main advantage to the formulation of \eqref{chi0} is that the two integrals may be computed directly using fast Fourier transforms of \(\psi_{n\textbf{k}}^*(\textbf{r})e^{-i\textbf{q}\cdot\textbf{r}}\psi_{n'\textbf{k}+\textbf{q}}(\textbf{r})\).  

Including local field effects, one may obtain the inverse macroscopic dielectric function \(\varepsilon^{-1}(\textbf{q},\omega)\) within the random phase approximation (RPA) as the solution of a Dyson's equation in terms of the non-interacting density-density response function $\chi_{\textbf{GG}'}^0(\textbf{q},\omega)$ of the form
\begin{eqnarray}
\varepsilon^{-1}(\textbf{q},\omega) &\approx& \left.\left[\delta_{\textbf{G}\textbf{G}'} - v_{\textbf{GG}'}(\textbf{q})\chi_{\textbf{GG}'}^0(\textbf{q},\omega)\right]^{-1}\right|_{\textbf{G} = \textbf{G}' = 0},\label{3}
\end{eqnarray}
where \(\delta_{\textbf{GG}'}\) is the Kronecker delta, and $v_{\textbf{G}}(\textbf{q})$ is the Fourier transform of the Coulomb kernel.  As \eqref{3} is a Dyson's equation, its solution includes many-body effects within linear response.  Note that the inclusion of exchange and correlation effects in $v_{\textbf{G}}(\textbf{q})$ at the LDA level adds a minor correction to the present results, as already shown for the case of graphite \cite{Pichler,splitting} and transition metal dichalcogenides \cite{KristenDichalcogenides2013}.

As discussed in Ref.~\cite{RadialCutoff}, for a 3D periodic system with translational invariance, the Coulomb kernel is
\begin{eqnarray}
v^{3\textrm{D}}_{\textbf{GG}'}(\textbf{q}) &=& \delta_{\textbf{GG}'}\iiint d\textbf{r}\frac{e^{i(\textbf{q}+\textbf{G})\cdot\textbf{r}}}{\|\textbf{r}\|} = \frac{4\pi}{\|\textbf{q}+\textbf{G}\|^2}\delta_{\textbf{GG}'}.
\label{culker}
\end{eqnarray}

However, for a system which is periodic in only two dimensions, such as a bulk slab or graphene, interactions between periodic images in a TDDFT-RPA calculation may be significant due to the long-range behaviour of $v^{3\textrm{D}}$.  This will be the case even for systems with sufficient vacuum to converge the electron density at the DFT level.  On the other hand, image--image interactions are included at the TDDFT-RPA level only through $v^{3\textrm{D}}$.  This motivates us to introduce a 2D periodic Coulomb kernel, $v^{2\textrm{D}}$, which is both translationally invariant and zero for $|z| > R$, where $R$ is the ``radial cutoff'' for the Coulomb kernel.  In this way, interactions between periodically repeated images are explicitly removed.

The 2D periodic Coulomb kernel of the radial cutoff method \cite{RadialCutoff} is then
\begin{eqnarray}
v^{2\textrm{D}}_{\textbf{GG}'}(\textbf{q}) &=& \delta_{\textbf{GG}'}\int_{-R}^R dz\iint dx dy \frac{e^{i (\textbf{q}+\textbf{G})\cdot(\textbf{x}+\textbf{y}+\textbf{z})}}{\sqrt{x^2+y^2+z^2}}\nonumber\\
 &=& \frac{4\pi}{\|\textbf{q}+\textbf{G}_\|\|} \delta_{\textbf{GG}'} \int_0^R \cos(G_z z) e^{-\|\textbf{q}+\textbf{G}_\|\|z} dz\nonumber\\
&=& \frac{4\pi\left[1 + e^{-\|\textbf{q}+\textbf{G}_\|\|R}\left[\frac{G_z \sin G_z R}{\|\textbf{q}+\textbf{G}_\|\|} - \cos G_zR\right]\right]}{\|\textbf{q}+\textbf{G}\|^2}\delta_{\textbf{GG}'}.
\label{rcfker}
\end{eqnarray}
Employing the suggested choice of $R = \frac{L_z}{2}$ from Ref.~\cite{RadialCutoff}, since $G_z = \frac{2\pi n_z}{L_z}$, where $n_z \in \mathbb{Z}$, one finds
\begin{eqnarray}
v^{2\textrm{D}}_{\textbf{GG}'}(\textbf{q}) &=& \frac{4\pi\left[1 - (-1)^{n_z} e^{-\|\textbf{q}+\textbf{G}_\|\|\frac{L_z}{2}}\right]}{\|\textbf{q}+\textbf{G}\|^2}\delta_{\textbf{GG}'}.\label{v2D}
\label{rcoff}
\end{eqnarray}
From Eqn.~\eqref{v2D} we clearly see that for $L_z \gg 2/q$ or $q \gtrsim$ 1~\AA$^{-1}$, $v^{2\textrm{D}} \rightarrow v^{3\textrm{D}}$.  

Note that by choosing $R = \frac{L_z}{2}$, interactions between densities within the cell, but within $\frac{L_z}{4}$ of the cell boundary in the $z$-direction, are artificially removed.  For graphene, where the electron density occupies a narrow region within the center of the cell, this does not pose major difficulties.  However, if one were to consider a thick slab, i.e., a surface, choosing $R = \frac{L_z}{2}$ would remove interactions between the two surfaces of the slab.  On the other hand, using a larger radial cutoff would include unwanted interactions between repeated images.  

As an alternative, one may introduce further regions of vacuum separating repeated images directly at the TDDFT-RPA level. Although unoccupied wave functions may be non-zero in the vacuum region, e.g., plane-waves, occupied wave functions, i.e., the density, are negligible.  For this reason, in the added vacuum regions the matrix elements for the occupied KS wavefunctions are always zero, and the inclusion of extra vacuum in Eqn.~\eqref{chi0} only enters into the non-interacting density-density response function through the unit cell volume $\Omega$, and hence the space over which the fast Fourier transforms of \(\psi_{n\textbf{k}}^*(\textbf{r})e^{-i\textbf{q}\cdot\textbf{r}}\psi_{n'\textbf{k}+\textbf{q}}(\textbf{r})\) are calculated, and the reciprocal unit cell vectors $\textbf{G}$.  

We may thus simply introduce extra unit cells of vacuum, or ``zero padding'' in the non-periodic direction, by doubling or tripling $L_z$ when computing the set of $\textbf{G}$ vectors to include at the TDDFT-RPA level.  
In this way, increasing the length of the unit cell in the non-periodic direction through the inclusion of vacuum effectively increases the density of sampling of the reciprocal unit cell.  However, the ``zero padding'' method only provides a computational advantage when performing the initial DFT calculation of the KS orbitals.  At the TDDFT-RPA level, the computational expense is unchanged.  Further, image--image interactions are still present with this method.

To solve the aforementioned difficulties with the radial cutoff and zero padding methods, one may combine both approaches.  By doubling the unit cell in the $z$-direction via zero padding, one may ensure all interactions within the unit cell are included, and spurious image--image interactions are removed, using a radial cutoff of $R = \frac{L_z}{2}$.  It is this combination of both methods which provides the most efficient means to describe bulk surfaces.

Finally, the quantities of fundamental interest are the  loss function \(-\Im\{\varepsilon^{-1}(\textbf{q},\omega)\}\), the absorption or imaginary part of the dielectric function \(\Im\{\varepsilon(\textbf{q},\omega)\}\), and the real part of the dielectric function \(\Re\{\varepsilon(\textbf{q},\omega)\}\), which may be obtained from Eqn.~\eqref{3}.  

\begin{figure}
\centering
\includegraphics[width=0.8125\columnwidth]{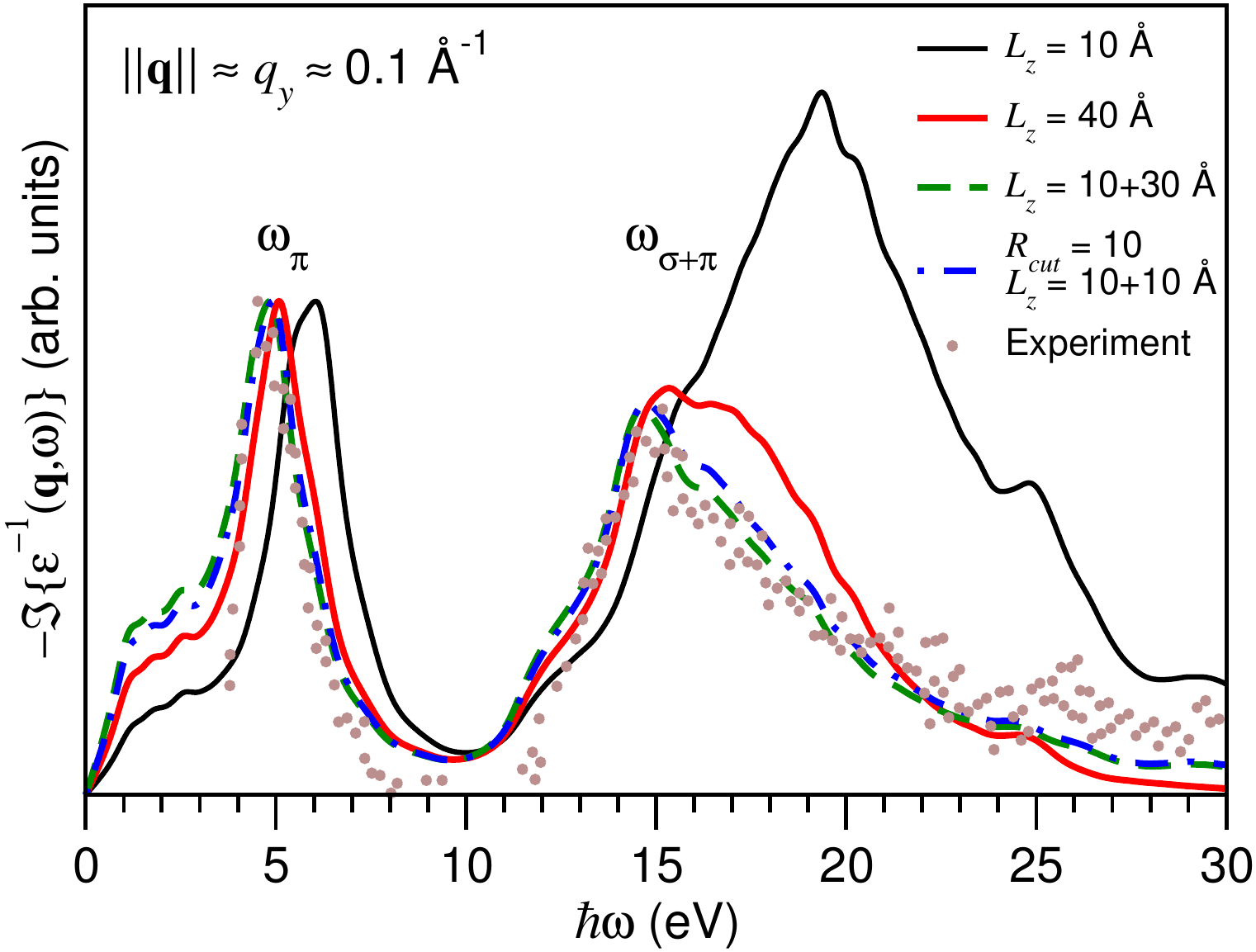}
\caption{Loss function $-\Im\{\varepsilon^{-1}(\textbf{q},\omega)\}$ versus energy $\hbar\omega$ in eV from TDDFT-RPA calculations for $\|\textbf{q}\|= q_y\approx 0.1$~\AA$^{-1}$ and $\gamma \approx 0.5$~eV, from standard DFT with $L_z \approx 10$~\AA\ (-----), $L_z \approx 40$~\AA\ ({\color{red}{\textbf{------}}}), augmented with zero-padding so $L_z \approx 10 + 30 \approx 40$~\AA\ (--~--~--), and including a radial cutoff so $L_z \approx 10 + 10 \approx 20$~\AA, $R \approx L/2 \approx 10$~\AA\ ({\color{blue}{\textbf{--~$\cdot$~--~$\cdot$}}}).  Experimental data from Ref.~\cite{exp1} ($\color{Tan}{\bullet}$) is provided for comparison.}\label{lossqy01}
\end{figure}

\begin{figure}
\centering
\includegraphics[width=0.8125\columnwidth]{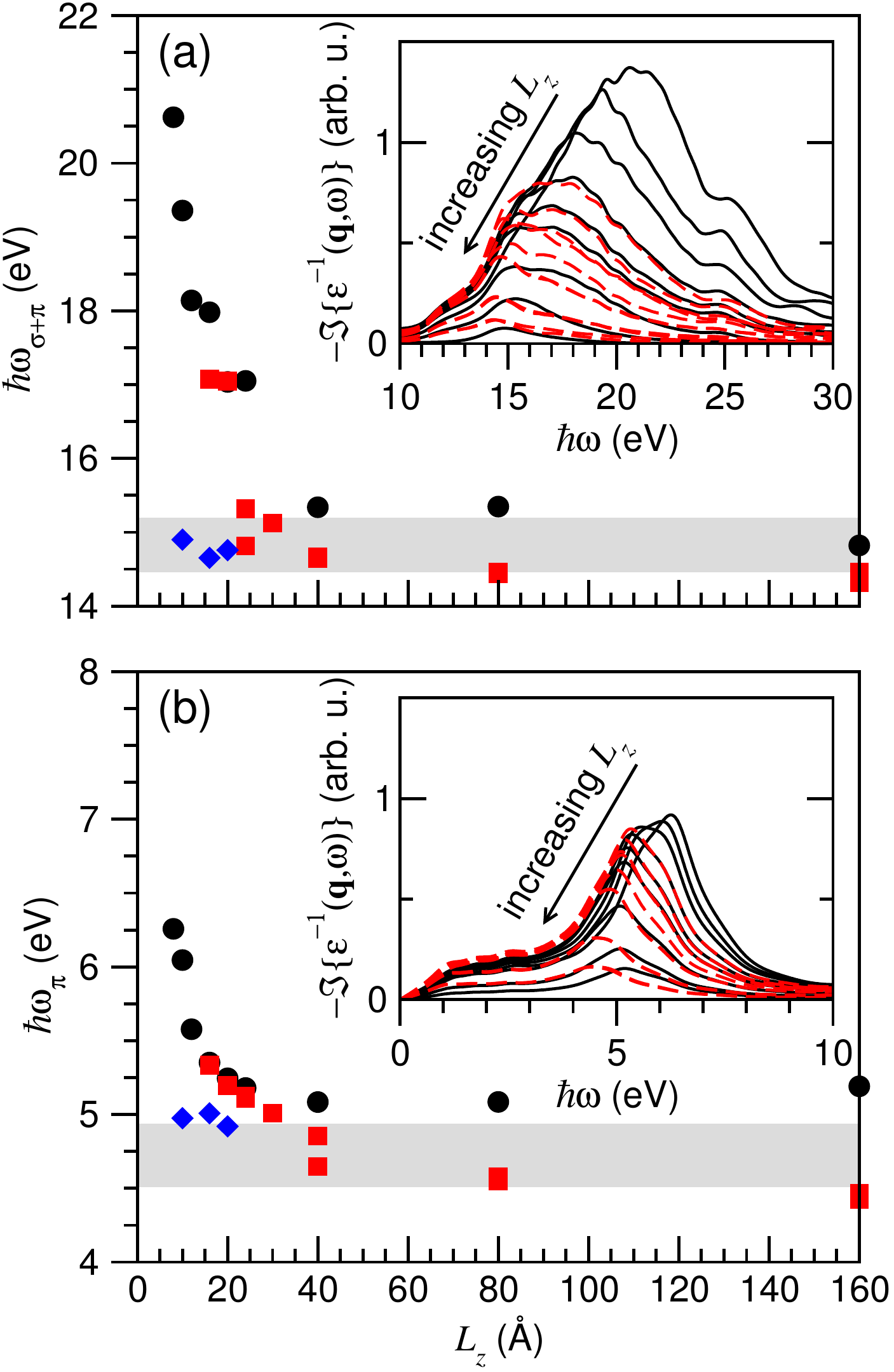}
\caption{Plasmon energies in eV versus unit cell parameter $L_z$ in \AA\ for the (a) $\sigma+\pi$ plasmon $\omega_{\sigma+\pi}$ and (b) $\pi$ plasmon $\omega_{\pi}$, obtained from TDDFT-RPA calculations of $\max_{\omega}-\Im\{\varepsilon^{-1}(\textbf{q},\omega)\}$ for $\|\textbf{q}\| = q_y \approx 0.1$~\AA$^{-1}$ (insets), from standard DFT ({$\medbullet$},\textbf{------}), augmented by zero-padding ({\color{red}{$\blacksquare$},\textbf{--~--~--}}), and including a radial cutoff of $R \approx L_z/2$ ({\color{blue}{$\Diamondblack$}}). Grey regions denote experimental range of $\omega_{\sigma+\pi}$ and $\omega_{\pi}$ \cite{exp1}.
}\label{omegavsL}
\end{figure}

\section{Results \& Discussion}\label{Results}

\begin{figure}
\centering
\includegraphics[width=0.985\columnwidth]{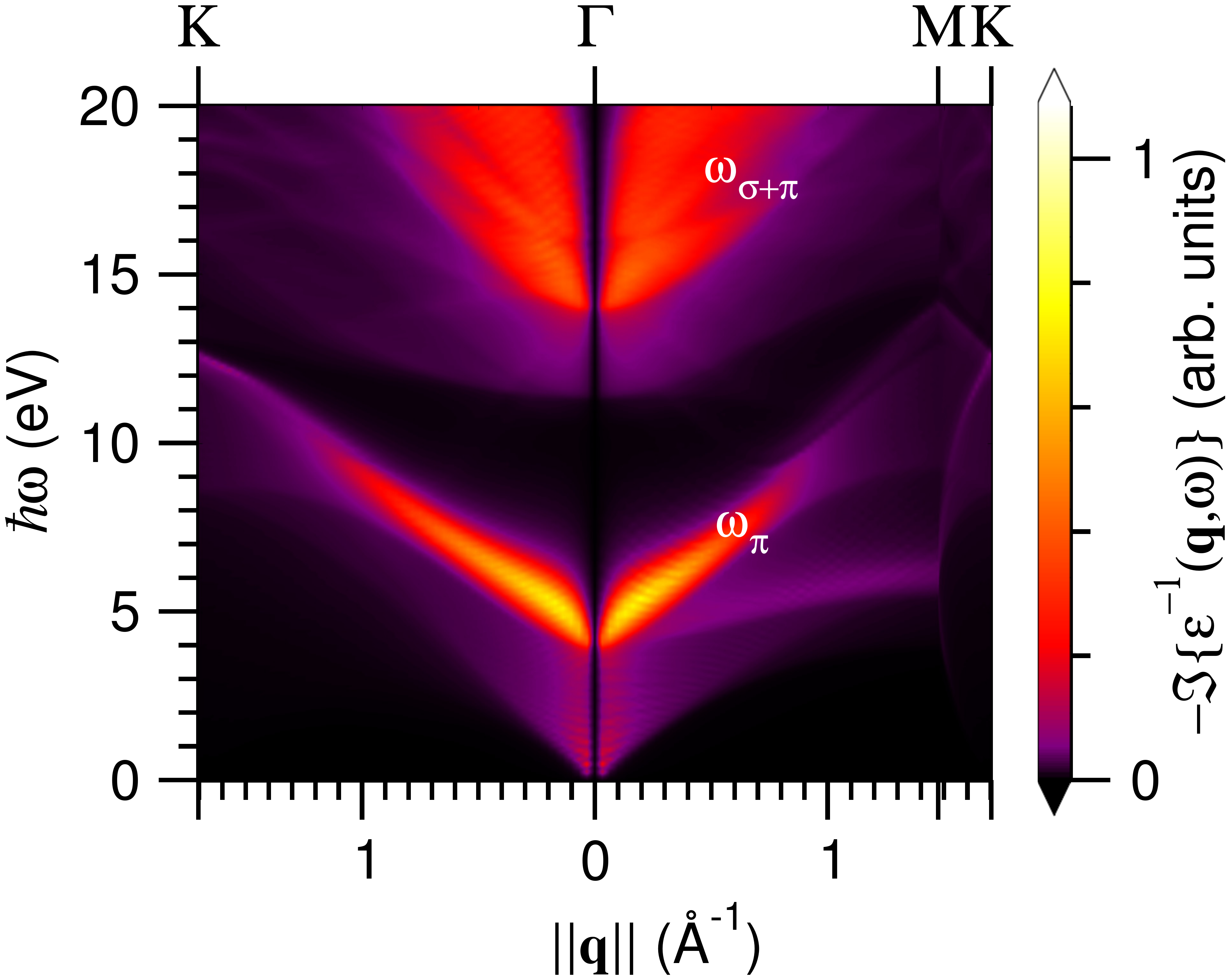}
\caption{Graphene loss function $-\Im\{\varepsilon^{-1}(\textbf{q},\omega)$ as a function of energy $\hbar\omega$ in eV and momentum transfer $\mathbf{q}$ in~\AA$^{-1}$ along the zigzag $\Gamma\rightarrow\mathrm{K}$ direction, armchair $\Gamma\rightarrow\mathrm{M}$ direction, and between $\Gamma\rightarrow\mathrm{M}$ and $\Gamma\rightarrow\mathrm{K}$, with $\gamma \approx 0.1$~eV.
}\label{LossDispersion}
\end{figure}

\begin{figure*}
\sidecaption
\centering
\includegraphics[width=1.487\columnwidth]{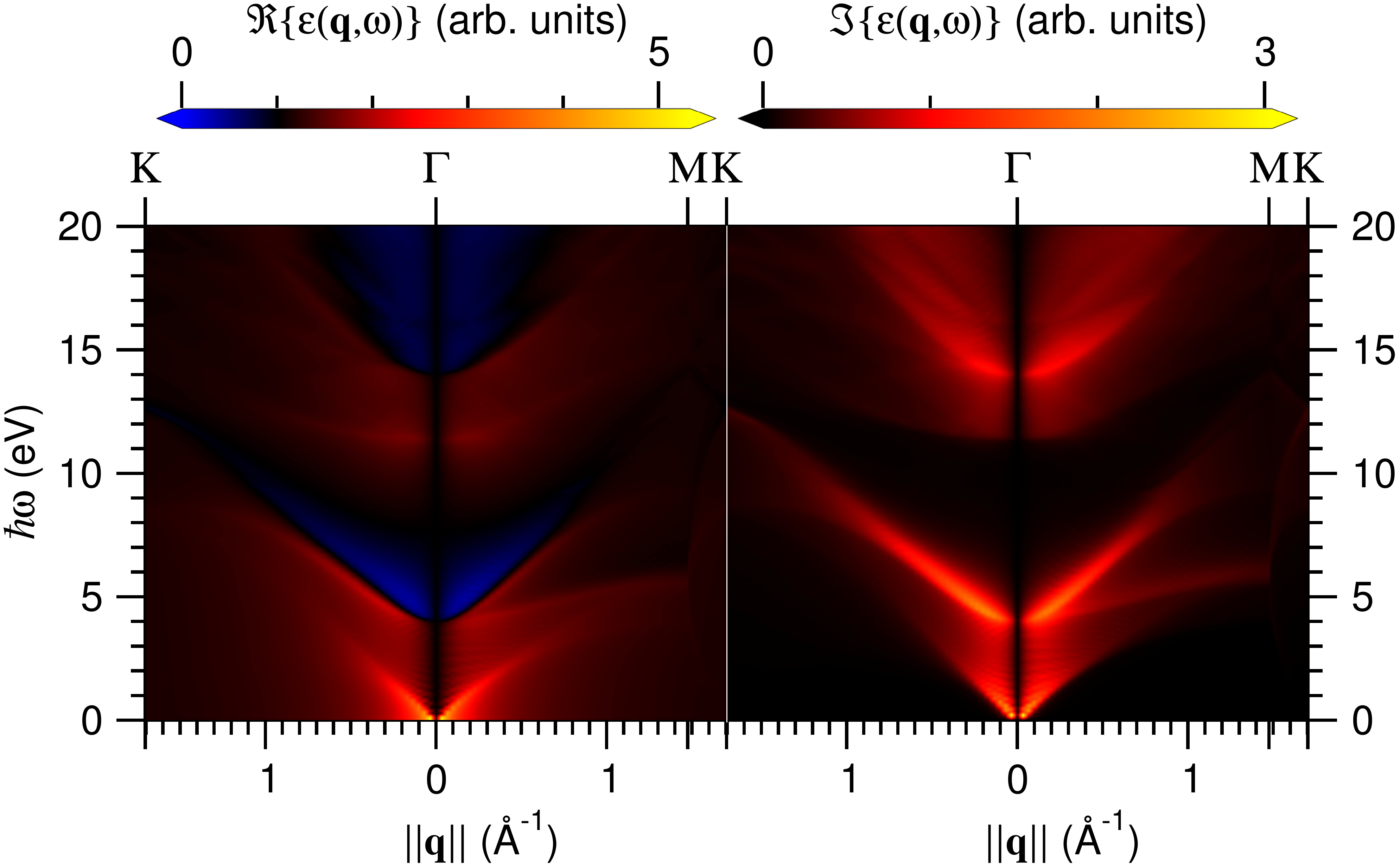}
\caption{Graphene macroscopic dielectric function (a) $\Re\{\varepsilon(\textbf{q},\omega)\}$ and absorption (b) $\Im\{\varepsilon(\textbf{q},\omega)\}$ as a function of energy $\hbar\omega$ in eV and momentum transfer $\textbf{q}$ in~\AA$^{-1}$ along the zigzag $\Gamma\rightarrow\mathrm{K}$ direction, armchair $\Gamma\rightarrow\mathrm{M}$ direction, and between $\Gamma\rightarrow\mathrm{M}$ and $\Gamma\rightarrow\mathrm{K}$.
}\label{ReImeps}
\end{figure*}

Figure \ref{lossqy01} shows the TDDFT-RPA calculated loss function at $q_y$ = 0.1~\AA$^{-1}$ at the various levels of approximation employed herein, relative to the measured loss function for graphene provided in Ref.~\cite{exp1}.  As shown in Fig.~\ref{FermiSurface}, a momentum transfer $q_y$ is along the $\Gamma\rightarrow M$ direction in reciprocal space.  To provide a clearer comparison between the various methods employed, all loss functions plotted in Fig.~\ref{lossqy01} have been normalized to have a consistent magnitude for the $\pi$ plasmon $\omega_\pi$.  

The TDDFT-RPA calculation based on a standard DFT calculation with $L_z \approx 10$~\AA\ has both the $\pi$ plasmon and $\sigma+\pi$ plasmon shifted to higher energies.  For $\omega_\pi$ this amounts to a shift of more than 1~eV relative to experiment.  On the other hand, for the $\sigma+\pi$ peak we see much higher energy ($\sim 4$~eV) transitions contributing to the loss, which have significantly higher intensities than those observed in the experiment.  We also see a lower energy shoulder in the main $\sigma+\pi$ peak, which is probably the relevant structure observed in the experimental loss function.  

When the vacuum layer at the DFT level is increased significantly to $L_z \approx 40$~\AA, both the $\omega_\pi$ and $\omega_{\sigma+\pi}$ plasmons are redshifted towards the experimental values, and the intensity of the $\omega_{\sigma+\pi}$ peak is reduced significantly.  However, even with such a large amount of vacuum included, the shape of the high energy $\sigma+\pi$ portion of the spectrum has more weight at higher energies ($\sim17$~eV) than the experimental spectrum.

If instead we include three unit cells of vacuum within the zero padding method, i.e., $L_z \approx 10 + 30 \approx 40$~\AA, the calculated and measured loss functions are in near quantitative agreement.  Both $\pi$ and $\sigma+\pi$ plasmons, their relative intensities, and the overall shape of the spectrum are very well reproduced.  Note that this calculation is based directly on the DFT calculation with $L_z \approx 10$~\AA, with vacuum being added only at the TDDFT-RPA level.  This clearly indicates that image--image interactions are responsible for the poor agreement obtained for TDDFT-RPA based on a standard DFT calculation with $L_z \approx 10$~\AA.  However, the computational expense of such a large zero padding calculation is not significantly reduced from a full DFT calculation with increased vacuum, as the TDDFT-RPA part is the bottle-neck in such calculations.  However, by combining both radial cutoff and zero padding methods, one obtains with only a single layer of zero padding ($L_z \approx 10 + 10 \approx 20$~\AA) an almost identical spectrum to that obtained employing three layers of zero padding.  This clearly indicates the equivalence of both methodologies.  

To provide a more quantitative comparison between TDDFT-RPA loss functions calculated with standard DFT, employing zero-padding, and combining with a radial cutoff, the calculated $\sigma+\pi$ and $\pi$ plasmons from each method are shown in Fig.~\ref{omegavsL} versus the amount of vacuum $L_z$ included at the TDDFT-RPA level.  The zero-padding plasmon energies agree resonably well with those based on standard DFT calculations with the same amount of vacuum. Moreover, the overall shape of the spectra for both methods, shown as insets in Fig.~\ref{omegavsL}, are quite similar.  Furthermore, while the zero-padding plasmon energies and spectra depend on $L_z$, they are independent of the initial amount of vacuum employed in the DFT calculations, whether 10 or 80~\AA. However, in all cases, convergence of the plasmon energies requires $L_z \gtrsim 30$~\AA, with the shape of the spectra continuing to change up to $L_z = 160$~\AA.  

If instead the radial cutoff method is employed, we see that even for only $L_z \approx 10$~\AA, the plasmon energies and shape of the spectrum are already converged with experiment.  These results clearly indicate the need for a radial cutoff of the Coulomb interaction to reproduce the measured loss function of isolated graphene with a reasonable computational effort.

To calculate the dispersion of the loss function, absorbance, and dielectric function, a denser \textbf{k}-point sampling, to yield a higher momentum transfer resolution $\Delta q \sim 0.03$~\AA$^{-1}$, has been employed, combining the zero padding ($L_z \approx 10 + 10 \approx 20$~\AA) and radial cutoff ($R \approx \frac{L_z}{2} \approx 10$~\AA) methodologies.  The dispersion of the TDDFT-RPA loss function for momentum transfer \textbf{q} along the high symmetry directions (see Fig.~\ref{FermiSurface}) is shown in Fig.~\ref{LossDispersion}.

Graphene's loss function consists primarily of a $\pi$ plasmon peak, which disperses quasi-linearly between 5 and 10~eV, and a broad $\sigma+\pi$ plasmon peak, which disperses between 14 and 20~eV, and broadens significantly with momentum transfer. Both $\pi$ and $\sigma+\pi$ plasmons have quite similar intensities and dispersions in both the $\Gamma\rightarrow \mathrm{K}$ and $\Gamma\rightarrow \mathrm{M}$ directions.  However, for momentum transfer parallel to the $\Gamma\rightarrow \mathrm{M}$ direction, there is also a more weakly dispersive peak below the $\pi$ plasmon.  This weakly dispersive peak is not seen for momentum transfers parallel to the $\Gamma \rightarrow \mathrm{K}$ direction.  In fact, this peak is already well described when including only a single unoccupied orbital in the TDDFT-RPA calculation.  This demonstrates this weakly dispersing peak is related to excitations to the $\pi^*$ band of graphene.

As shown in Fig.~\ref{FermiSurface}, the bonding $\pi$ and antibonding $\pi^*$ bands of graphene are both quite flat along the $\mathrm{M}\rightarrow\mathrm{M}'$ direction.  In fact, the tight-binding band structure is completely flat between $\mathrm{M}$ and $\mathrm{M}'$.  Momentum transfers parallel to the $\Gamma\rightarrow \mathrm{K}$ direction, cannot be from $\mathrm{M}\rightarrow\mathrm{M}'$.  Altogether this explains the observed lack of a weakly dispersing peak in the loss function for momentum transfer parallel to the $\Gamma\rightarrow\mathrm{K}$ direction in Fig.~\ref{LossDispersion}.  

Figure~\ref{ReImeps} shows the dispersion of the similarly calculated TDDFT-RPA real and imaginary parts of the dielectric function.  The blue regions shown in Fig.~\ref{ReImeps}(a) demonstrate that the observed $\omega_\pi$ and $\omega_{\sigma+\pi}$ peaks in the loss function are due to plasmons.  However, the weakly dispersing peak observed for momentum transfer parallel to the $\Gamma\rightarrow\mathrm{M}$ direction is related to a peak in the imaginary part of the dielectric function, rather than a plasmon mode.

\section{Conclusions}\label{Conclusions}

The TDDFT-RPA implementation within \textsc{gpaw} has been extended to employ both a radial cutoff of the Coulomb kernel $v^{2\textrm{D}}$ for 2D periodic systems, and include zero padding via extra unit cells of vacuum at the TDDFT-RPA level.
The spurious image---image interactions have a significant impact on the calculated loss function for isolated systems, and must be removed to describe the measured loss function correctly, as demonstrated for graphene.  These results are particularly important in the area of nanoplasmonics, and for the description of the low energy free-charge carrier plasmons induced by electrostatic or potassium doping.  

\begin{acknowledgement}
The author thanks V.\ Despoja, L.\ N.\ Glanzmann, C.\ Kramberger, P.\ Ayala, T.\ Pichler, and A. Rubio for fruitful discussions, and funding through the Spanish Grants (FIS2010-21282-C02-01) and (PIB2010US-00652), ``Grupos Consolidados UPV/EHU del Gobierno Vasco'' (IT-578-13) and the Spanish ``Juan de la Cierva'' program (JCI-2010-08156).
  \end{acknowledgement}

\bibliographystyle{pss}
\bibliography{bibliography}

\providecommand{\WileyBibTextsc}{}
\let\textsc\WileyBibTextsc
\providecommand{\othercit}{}
\providecommand{\jr}[1]{#1}
\providecommand{\etal}{~et~al.}


\begin{thebibliography}{[10]}

\bibitem{novoselov}
 \textsc{K.\,S. Novoselov},  \textsc{A.\,K. Geim},  \textsc{S.\,V. Morozov},
  \textsc{D.~Jiang},  \textsc{Y.~Zhang},  \textsc{S.\,V. Dubonos},
  \textsc{I.\,V. Grigorieva},  and  \textsc{A.\,A. Firsov},
 \jr{Science} \textbf{306}(5696), 666--669 (2004).


\bibitem{Novoselov05N}
 \textsc{K.\,S. Novoselov},  \textsc{A.\,K. Geim},  \textsc{S.\,V. Morozov},
  \textsc{D.~Jiang},  \textsc{M.\,I. Katsnelson},  \textsc{I.\,V. Grigorieva},
  \textsc{S.\,V. Dubonos},  and  \textsc{A.\,A. Firsov},
 \jr{Nature} \textbf{438}(7065), 197--200 (2005).


\bibitem{Geim07NM}
 \textsc{A.\,K. Geim} and  \textsc{K.\,S. Novoselov},
 \jr{Nat. Mater.} \textbf{6}(3), 183--191 (2007).


\bibitem{CastroNeto09RMP}
 \textsc{A.\,H. Castro~Neto},  \textsc{F.~Guinea},  \textsc{N.\,M.\,R. Peres},
  \textsc{K.\,S. Novoselov},  and  \textsc{A.\,K. Geim},
 \jr{Rev. Mod. Phys.} \textbf{81}(1), 109--162 (2009).


\bibitem{Novodsel1}
 \textsc{K.\,S. Novoselov},  \textsc{A.\,K. Geim},  \textsc{S.\,V. Morozov},
  \textsc{D.~Jiang},  \textsc{M.\,I. Katsnelson},  \textsc{I.\,V. Grigorieva},
  \textsc{S.\,V. Dubonos},  and  \textsc{A.\,A. Firsov},
 \jr{Nature} \textbf{438}, 197--200 (2005).


\bibitem{Novodsel2}
 \textsc{Y.~Zhang},  \textsc{Y.\,W. Tan},  \textsc{H.\,L. Stormer},  and
  \textsc{P.~Kim},
 \jr{Nature} \textbf{438}, 201--204 (2005).


\bibitem{Novodsel3}
 \textsc{K.\,S. Novoselov},  \textsc{A.\,K. Geim},  \textsc{S.\,V. Morozov},
  \textsc{D.~Jiang},  \textsc{Y.~Zhang},  \textsc{S.\,V. Dubonos},
  \textsc{I.\,V. Grigorieva},  and  \textsc{A.\,A. Firsov},
 \jr{Science} \textbf{306}, 666--669 (2004).


\bibitem{Novodsel4}
 \textsc{K.\,S. Novoselov},  \textsc{D.~Jiang},  \textsc{F.~Schedin},
  \textsc{T.\,J. Booth},  \textsc{V.\,V. Khotkevich},  \textsc{S.\,V. Morozov},
   and  \textsc{A.\,K. Geim},
 \jr{Proc. Natl. Acad. Sci. U.S.A.} \textbf{102}(30), 10451--10453 (2005).


\bibitem{Novodsel6}
 \textsc{A.\,K. Geim} and  \textsc{K.\,S. Novoselov},
 \jr{Nature Mater.} \textbf{6}, 183--191 (2007).


\bibitem{Pletikos}
 \textsc{I.~Pletikosi{\'{c}}},  \textsc{M.~Kralj},  \textsc{P.~Pervan},
  \textsc{R.~Brako},  \textsc{J.~Coraux},  \textsc{A.\,T. N'Diaye},
  \textsc{C.~Busse},  and  \textsc{T.~Michely},
 \jr{Phys. Rev. Lett.} \textbf{102}, 056808 (2009).


\bibitem{RevModPhys.83.1193}
 \textsc{M.\,O. Goerbig},
 \jr{Rev. Mod. Phys.} \textbf{83}, 1193--1243 (2011).


\bibitem{RevModPhys.83.407}
 \textsc{S.~Das~Sarma},  \textsc{S.~Adam},  \textsc{E.\,H. Hwang},  and
  \textsc{E.~Rossi},
 \jr{Rev. Mod. Phys.} \textbf{83}, 407--470 (2011).


\bibitem{RevModPhys.81.109}
 \textsc{A.\,H. Castro~Neto},  \textsc{F.~Guinea},  \textsc{N.\,M.\,R. Peres},
  \textsc{K.\,S. Novoselov},  and  \textsc{A.\,K. Geim},
 \jr{Rev. Mod. Phys.} \textbf{81}, 109--162 (2009).


\bibitem{0034-4885-74-8-082501}
 \textsc{A.\,H.\,C. Neto} and  \textsc{K.~Novoselov},
 \jr{Rep. Prog. Phys.} \textbf{74}(8), 082501 (2011).


\bibitem{TDDFTGraphene}
 \textsc{V.~Despoja},  \textsc{D.\,J. Mowbray},
  \textsc{D.~Vlahovi\ifmmode\,\acute{c}\else \'{c}\fi{}},  and
  \textsc{L.~Maru\ifmmode\,\check{s}\else\,\v{s}\fi{}i\ifmmode\,\acute{c}\else
  \'{c}\fi{}},
 \jr{Phys. Rev. B} \textbf{86}, 195429 (2012).


\bibitem{GrapheneExciton}
 \textsc{V.~Despoja},  \textsc{I.~Lon\v{c}ari\'{c}},  \textsc{D.\,J. Mowbray},
  and  \textsc{L.~Maru\v{s}i\'{c}},
 \jr{Phys. Rev. B} \textbf{88}, 235437 (2013).


\bibitem{Pichler}
 \textsc{A.\,G. Marinopoulos},  \textsc{L.~Reining},  \textsc{V.~Olevano},
  \textsc{A.~Rubio},  \textsc{T.~Pichler},  \textsc{X.~Liu},
  \textsc{M.~Knupfer},  and  \textsc{J.~Fink},
 \jr{Phys. Rev. Lett.} \textbf{89}(7), 076402 (2002).


\bibitem{EELS}
 \textsc{A.\,G. Marinopoulos},  \textsc{L.~Reining},  \textsc{A.~Rubio},  and
  \textsc{V.~Olevano},
 \jr{Phys. Rev. B} \textbf{69}, 245419 (2004).


\bibitem{splitting}
 \textsc{C.~Kramberger},  \textsc{R.~Hambach},  \textsc{C.~Giorgetti},
  \textsc{M.\,H. R\"ummeli},  \textsc{M.~Knupfer},  \textsc{J.~Fink},
  \textsc{B.~B\"uchner},  \textsc{L.~Reining},  \textsc{E.~Einarsson},
  \textsc{S.~Maruyama},  \textsc{F.~Sottile},  \textsc{K.~Hannewald},
  \textsc{V.~Olevano},  \textsc{A.\,G. Marinopoulos},  and
  \textsc{T.~Pichler},
 \jr{Phys. Rev. Lett.} \textbf{100}(19), 196803 (2008).


\bibitem{Marinopoulos02PRL}
 \textsc{A.\,G. Marinopoulos},  \textsc{L.~Reining},  \textsc{V.~Olevano},
  \textsc{A.~Rubio},  \textsc{T.~Pichler},  \textsc{X.~Liu},
  \textsc{M.~Knupfer},  and  \textsc{J.~Fink},
 \jr{Phys. Rev. Lett.} \textbf{89}(7), 076402 (2002).


\bibitem{Kramberger08PRL}
 \textsc{C.~Kramberger},  \textsc{R.~Hambach},  \textsc{C.~Giorgetti},
  \textsc{M.\,H. R\"{u}mmeli},  \textsc{M.~Knupfer},  \textsc{J.~Fink},
  \textsc{B.~B\"{u}chner},  \textsc{L.~Reining},  \textsc{E.~Einarsson},
  \textsc{S.~Maruyama},  \textsc{F.~Sottile},  \textsc{K.~Hannewald},
  \textsc{V.~Olevano},  \textsc{A.\,G. Marinopoulos},  and
  \textsc{T.~Pichler},
 \jr{Phys. Rev. Lett.} \textbf{100}, 196803 (2008).


\bibitem{MiskoSpe}
 \textsc{V.\,B. Jovanovi{\'{c}}},  \textsc{I.~Radovi{\'{c}}},
  \textsc{D.~Borka},  and  \textsc{Z.\,L. Mi{\v{s}}kovi{\'{c}}},
 \jr{Phys. Rev. B} \textbf{84}, 155416 (2011).


\bibitem{DespojaGraphene}
 \textsc{V.~Despoja},  \textsc{K.~Dekani\ifmmode\,\acute{c}\else \'{c}\fi{}},
  \textsc{M.~\ifmmode\,\check{S}\else\,\v{S}\fi{}unji\ifmmode\,\acute{c}\else
  \'{c}\fi{}},  and
  \textsc{L.~Maru\ifmmode\,\check{s}\else\,\v{s}\fi{}i\ifmmode\,\acute{c}\else
  \'{c}\fi{}},
 \jr{Phys. Rev. B} \textbf{86}, 165419 (2012).


\bibitem{response1}
 \textsc{J.~Yan},  \textsc{K.\,S. Thygesen},  and  \textsc{K.\,W.
  Jacobsen},
 \jr{Phys. Rev. Lett.} \textbf{106}, 146803 (2011).


\bibitem{response2}
 \textsc{J.~Yan},  \textsc{J.\,J. Mortensen},  \textsc{K.\,W. Jacobsen},  and
  \textsc{K.\,S. Thygesen},
 \jr{Phys. Rev. B} \textbf{83}, 245122 (2011).


\bibitem{exp1}
 \textsc{T.~Eberlein},  \textsc{U.~Bangert},  \textsc{R.\,R. Nair},
  \textsc{R.~Jones},  \textsc{M.~Gass},  \textsc{A.\,L. Bleloch},
  \textsc{K.\,S. Novoselov},  \textsc{A.~Geim},  and  \textsc{P.\,R.
  Briddon},
 \jr{Phys. Rev. B} \textbf{77}, 233406 (2008).


\bibitem{2D-exp}
 \textsc{C.~Tegenkamp},  \textsc{H.~Pfnur},  \textsc{T.~Langer},
  \textsc{J.~Baringhaus},  and  \textsc{H.\,W. Schumacher},
 \jr{J. Phys.: Condens. Matter} \textbf{23}, 012001 (2011).


\bibitem{exp2}
 \textsc{J.~Lu},  \textsc{K.\,P. Loh},  \textsc{H.~Huang},  \textsc{W.~Chen},
  and  \textsc{A.\,T.\,S. Wee},
 \jr{Phys. Rev. B} \textbf{80}, 113410 (2009).


\bibitem{KrambergerGrapheneEELS}
 \textsc{M.\,K. Kinyanjui},  \textsc{C.~Kramberger},  \textsc{T.~Pichler},
  \textsc{J.\,C. Meyer},  \textsc{P.~Wachsmuth},  \textsc{G.~Benner},  and
  \textsc{U.~Kaiser},
 \jr{Eur. Phys. Lett.} \textbf{97}(5), 57005 (2012).


\bibitem{RadialCutoff}
 \textsc{C.\,A. Rozzi},  \textsc{D.~Varsano},  \textsc{A.~Marini},
  \textsc{E.\,K.\,U. Gross},  and  \textsc{A.~Rubio},
 \jr{Phys. Rev. B} \textbf{73}(20), 205119 (2006).


\bibitem{KristenDichalcogenides2013}
 \textsc{K.~Andersen} and  \textsc{K.\,S. Thygesen},
 \jr{Phys. Rev. B} \textbf{88}, 155128 (2013).


\bibitem{GPAW}
 \textsc{J.\,J. Mortensen},  \textsc{L.\,B. Hansen},  and  \textsc{K.\,W.
  Jacobsen},
 \jr{Phys. Rev. B} \textbf{71}(3), 035109 (2005).


\bibitem{GPAWRev}
 \textsc{J.~Enkovaara},  \textsc{C.~Rostgaard},  \textsc{J.\,J. Mortensen},
  \textsc{J.~Chen},  \textsc{M.~Du{\l}ak},  \textsc{L.~Ferrighi},
  \textsc{J.~Gavnholt},  \textsc{C.~Glinsvad},  \textsc{V.~Haikola},
  \textsc{H.\,A. Hansen},  \textsc{H.\,H. Kristoffersen},  \textsc{M.~Kuisma},
  \textsc{A.\,H. Larsen},  \textsc{L.~Lehtovaara},  \textsc{M.~Ljungberg},
  \textsc{O.~Lopez-Acevedo},  \textsc{P.\,G. Moses},  \textsc{J.~Ojanen},
  \textsc{T.~Olsen},  \textsc{V.~Petzold},  \textsc{N.\,A. Romero},
  \textsc{J.~Stausholm-M{\o}ller},  \textsc{M.~Strange},  \textsc{G.\,A.
  Tritsaris},  \textsc{M.~Vanin},  \textsc{M.~Walter},  \textsc{B.~Hammer},
  \textsc{H.~H{\"{a}}kkinen},  \textsc{G.\,K.\,H. Madsen},  \textsc{R.\,M.
  Nieminen},  \textsc{J.\,K. N{\o}rskov},  \textsc{M.~Puska},  \textsc{T.\,T.
  Rantala},  \textsc{J.~Schi{\o}tz},  \textsc{K.\,S. Thygesen},  and
  \textsc{K.\,W. Jacobsen},
 \jr{J. Phys.: Condens. Matter} \textbf{22}(25), 253202 (2010).


\bibitem{LDA}
 \textsc{J.\,P. Perdew} and  \textsc{A.~Zunger},
 \jr{Phys. Rev. B} \textbf{23}, 5048--5079 (1981).


\end{thebibliography}
\end{document}